\documentclass[twocolumn]{aastex631}

\usepackage{natbib}
\usepackage[version=4]{mhchem}

\newcommand{\RNum}[1]{\uppercase\expandafter{\romannumeral #1\relax}}

\DeclareSymbolFont{UPM}{U}{eur}{m}{n}
\DeclareMathSymbol{\umu}{0}{UPM}{"16}
\let\oldumu=\umu
\renewcommand\umu{\ifmmode\oldumu\else\math{\oldumu}\fi}
\newcommand\micro{\umu}

\let\microns \jmicron

\def\ms{$\mathrm{m}^{-2}$}

\shorttitle{Constraining \ce{H2O} using Grids}
\shortauthors{Latouf et al.}

\begin{document}



\title{Bayesian Analysis for Remote Biosignature Identification on exoEarths (BARBIE) \RNum{1}: Using Grid-Based Nested Sampling in Coronagraphy Observation Simulations for \ce{H2O}}

\author[0000-0001-8079-1882]{Natasha Latouf}
\altaffiliation{NSF Graduate Research Fellow, 2415 Eisenhower Ave, Alexandria, VA 22314}
\affiliation{Department of Physics and Astronomy, George Mason University, 4400 University Drive MS 3F3, Fairfax, VA, 22030, USA}
\affiliation{NASA Goddard Space Flight Center, 8800 Greenbelt Road, Greenbelt, MD 20771, USA}
\affiliation{Sellers Exoplanents Environment Collaboration, 8800 Greenbelt Road, Greenbelt, MD 20771, USA}

\author[0000-0002-8119-3355]{Avi M. Mandell}
\affiliation{NASA Goddard Space Flight Center, 8800 Greenbelt Road, Greenbelt, MD 20771, USA}
\affiliation{Sellers Exoplanents Environment Collaboration, 8800 Greenbelt Road, Greenbelt, MD 20771, USA}

\author[0000-0002-2662-5776]{Geronimo L. Villanueva}
\affiliation{NASA Goddard Space Flight Center, 8800 Greenbelt Road, Greenbelt, MD 20771, USA}
\affiliation{Sellers Exoplanents Environment Collaboration, 8800 Greenbelt Road, Greenbelt, MD 20771, USA}

\author[0000-0001-7912-6519]{Michael Dane Moore}
\affiliation{NASA Goddard Space Flight Center, Greenbelt, MD, USA.}
\affiliation{Business Integra, Inc., Bethesda, MD, USA.}
\affiliation{Sellers Exoplanents Environment Collaboration, 8800 Greenbelt Road, Greenbelt, MD 20771, USA}

\author{Nicholas Susemiehl}
\affiliation{Center for Research and Exploration in Space Science and Technology, NASA Goddard Space Flight Center, Greenbelt, MD, USA.}
\affiliation{NASA Goddard Space Flight Center, Greenbelt, MD, USA.}

\author[0000-0002-5060-1993]{Vincent Kofman}
\affiliation{NASA Goddard Space Flight Center, 8800 Greenbelt Road, Greenbelt, MD 20771, USA}
\affiliation{Sellers Exoplanents Environment Collaboration, 8800 Greenbelt Road, Greenbelt, MD 20771, USA}
\affiliation{Integrated Space Science and Technology Institute, Department of Physics, American University, Washington DC}

\author[0000-0002-9338-8600]{Michael D. Himes}
\affiliation{NASA Postdoctoral Program Fellow, NASA Goddard Space Flight Center, 8800 Greenbelt Road, Greenbelt, MD 20771, USA}

\correspondingauthor{Natasha Latouf}
\email{nlatouf@gmu.edu, natasha.m.latouf@nasa.gov}

\begin{abstract}

Detecting \ce{H2O} in exoplanet atmospheres is the first step on the path to determining planet habitability. Coronagraphic design currently limits the observing strategy used to detect \ce{H2O}, requiring the choice of specific bandpasses to optimize abundance constraints. In order to examine the optimal observing strategy for initial characterization of habitable planets using coronagraph-based direct imaging, we quantify the detectability of \ce{H2O} as a function of signal-to-noise ratio (SNR) and molecular abundance across 25 bandpasses in the visible wavelength range (0.5-1 {\microns}). We use a pre-constructed grid consisting of 1.4 million geometric albedo spectra across a range of abundance and pressure, and interpolate to produce forward models for an efficient nested sampling routine, PSGnest. We first test the detectability of \ce{H2O} in atmospheres that mimic a modern-Earth twin, and then expand to examine a wider range of \ce{H2O} abundances; for each abundance value, we constrain the optimal 20\% bandpasses based on the effective signal-to-noise ratio (SNR) of the data. We present our findings of \ce{H2O} detectability as functions of SNR, wavelength, and abundance, and discuss how to use these results for optimizing future coronographic instrument design. We find that there are specific points in wavelength where \ce{H2O} can be detected down to 0.74 {\microns} with moderate-SNR data for abundances at the upper end of Earth's presumed historical values, while at 0.9 {\microns}, detectability is possible with low-SNR data at modern Earth abundances of \ce{H2O}.

\end{abstract}

\section{Introduction}
\label{sec:intro}

\begin{figure*}[]
\centering
\includegraphics[width=0.7\textwidth]{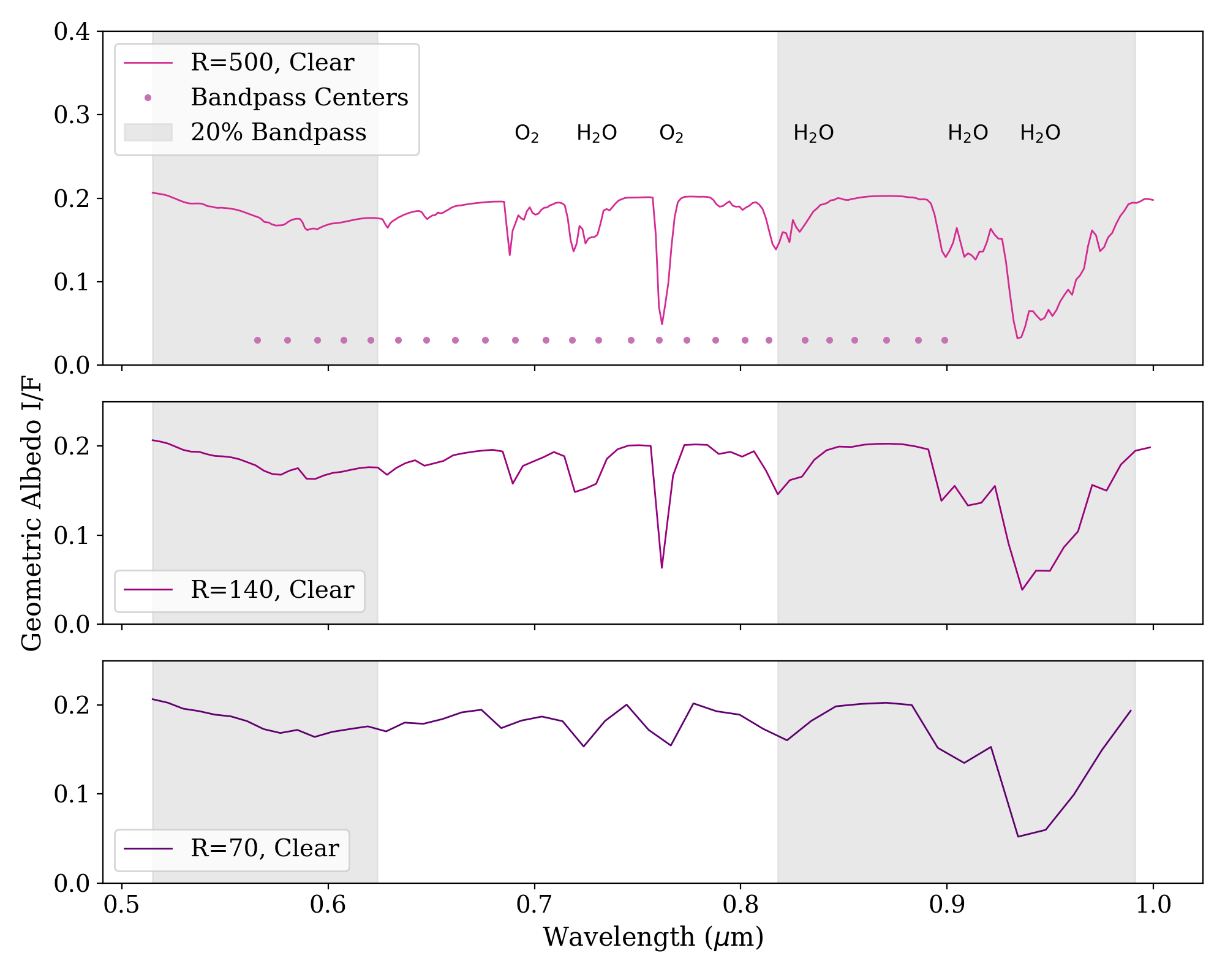}
\caption{The full modern-Earth high-resolution spectrum from the spectral grid in \citet{susemiehl23}, with a zero cloud fraction. We show the full resolution (R) of 500 (shown in pink, top panel), the spectrum binned to R=140 that are used in the nested sampling simulations conducted in this work,  (shown in purple, middle panel), and a spectrum at R=70 for reference (shown in dark purple, bottom panel). We also present the bandpass centers used in this study as dots (only in the top panel), and the first and last 20\% bandpasses shaded in grey. We note that bandpasses overlap heavily with each other, so we only show the first and last for clarity.}
\label{fig:spectra_intro}
\end{figure*} 

We have seen the discovery and confirmation of thousands of exoplanets since the first planet found orbiting a Sun-like star almost 30 years ago \citep{mayor95}, and we are now on the verge of entering an exciting new era of exoplanetary exploration: detection and characterization of terrestrial exoplanet atmospheres. Characterization of an exoplanet's atmosphere can provide vital information on the formation, history, and physical composition of the planet; for potentially habitable planets, we can also search their atmospheres for biosignatures such as \ce{H2O}, \ce{O2}, and \ce{O3} that can hint at the likelihood of clement conditions and biological activity \citep{kaltenegger17,schwieterman18}. With current advances in instrumentation and proposed future missions that are set to prioritize terrestrial exoplanet atmosphere characterization \citep[e.g.,][]{roberge18, luvoir}, the ability to find a habitable Earth-twin is becoming a realistic prospect within the next few decades. The looming goal is to discover, characterize, and analyze an Earth-twin terrestrial planet which could provide insight into our own Earth's history and formation and potentially evidence of biological activity. 
 
A true Earth-twin analog would occupy the habitable zone (HZ) of its host star at such a distance that liquid water can exist at the surface \citep{kasting93, kopparapu13}; hence, detecting \ce{H2O} generally stands as the first step in the search for habitability. Terrestrial planets in the HZ with strong water detections are the prime habitable planet candidates and thus will be priority targets for re-observation. However, observed Earth twins could exhibit a range of \ce{H2O} abundances; at different times in the Earth's history, the planet hosted varying concentrations of different atmospheric biomarkers. Interestingly, little geologic evidence of the atmospheric \ce{H2O} abundance through Earth's history is available. Even though other biomarkers such as \ce{O2} and \ce{O3} have geochemical proxies that allow for better estimation of varying epoch atmospheric abundances \citep{planavsky14}, the presence of \ce{H2O} in geochemical records does not constrain atmospheric abundance. This is due to the fact that water is present throughout the Earth's history and, unlike \ce{O2} and \ce{O3}, present in solid, liquid as well as gaseous phases. However, general circulation models (GCMs) have been able to provide constraints on Earth's water vapor abundance through time using constrained parameter sets.

\begin{deluxetable*}{cccc} 
    \tablewidth{0pt}
    \tablehead{
    \colhead{Parameter Symbol} & \colhead{Fiducial} & \colhead{Minimum} & \colhead{Maximum}
    }
    \startdata
        $\mathrm{C_f}$ & 0.5 & N/A & N/A\\
        \ce{H2O} & 3$\times10^{-3}$ VMR & $10^{-8}$ VMR & $10^{-1}$ VMR\\
        \ce{O3} & 7$\times10^{-7}$ VMR & $10^{-10}$ VMR & $10^{-1}$ VMR \\
        \ce{O2} & 0.21 VMR & $10^{-8}$ VMR & 0.8 VMR\\
        $\mathrm{P_0}$ & 1.0 Bar & $10^{-3}$ Bar & 10 Bar\\
        $\mathrm{g}$ & 9.8 m/s & 1 m/s & 100 m/s\\
        $\mathrm{A_s}$ & 0.3 & $10^{-2}$ & 1
    \enddata
    \caption{Modern Earth-like fiducial parameter values used in these simulations, and in S23, along with minimum and maximum values.}
    \label{tab:params}
\end{deluxetable*}



Looking forward to next-generation instrument design, such as the instrumentation for the upcoming Habitable Worlds Observatory (HWO), efficiency is key to maximizing the science output from limited observation time. With this in mind, the optimal wavelength for the spectral bandpass used for measuring the reflected flux and atmospheric absorption is a crucial factor to consider. Coronagraph designs for future exoplanet direct imaging mission concepts generally feature two deformable mirrors (DMs) to control the wavefront of light, along with a suite of focal plane masks (FPM) such as the vortex coronagraph and the apodized pupil Lyot designs, both of which were included for consideration in the LUVOIR and HabEx studies \citep{ruane15, luvoir, por20, roser22}. These designs have an out-of-band wavefront sensor that takes reflected light from the FPMs to sense pointing errors and wavefront drifts, but the accuracy of the wavelength control degrades as the observed wavelength diverges from the central wavelength. It is for this reason that when using a coronagraph, the light must pass through a bandpass filter, which filters out light beyond a defined range on either side of the central wavelength; light outside of this range may not be well corrected by the optical system and thus degrade overall performance. Currently, the width of bandpasses (i.e.~10, 20 or 30\% of the central wavelength) is an area of active technology development, but we chose to adopt an intermediate value of 20\% in our simulations.
 
There are multiple factors which affect the efficiency of atmospheric characterization at different wavelengths. First, when we observe at shorter wavelengths, we can observe the HZs of more distant and lower mass stars, as the HZs for these targets are smaller in angular separation and the inner working angle (IWA) of coronagraphs is proportional to wavelength. Second, prioritizing shorter wavelengths also enables us to decrease astrophysical noise terms, since the telescope point spread function (PSF) area is also proportional to $\lambda$, and background noise scales with PSF solid angle, thus allowing for less noisy observations. Third, the stellar SED peaks at 0.5 - 0.7 {\microns} for FGK-type stars; thus, positioning our bandpass effectively allows for the observation of a higher photon flux, thus increasing the achievable SNR with the same exposure time. However, \ce{H2O} has an increasing number of deep spectral features at longer wavelengths; thus, observing at longer wavelengths would allow us to minimize the SNR needed to detect these features but would decrease the photon flux and HZ visibility, while observing at shorter wavelengths with shorter features would optimize observing efficiency but would require a higher SNR. Quantifying the required SNR for \ce{H2O} detection in each wavelength region is therefore a crucial component to determining the optimal observing strategy. 

In order to quantify the SNR required to detect \ce{H2O} for different spectral bandpasses and different molecular abundances, we utilized a Bayesian spectral retrieval methodology. Spectral retrieval studies have been used to explore the detectability of atmospheric compositions for the direct imaging of exoplanets using both the Roman Space Telescope \citep[e.g.,][]{lupu16, nayak17} as well as a future Earth-twin imaging mission similar to HWO \citep{feng18, smith20, robinson22, damiano22}. However, traditional Bayesian retrievals using real-time radiative transfer calculations require large amounts of computational runtime to effectively explore multiple parameters and mission capabilities - and the ongoing and future addition of more complex model features such as high-accuracy aerosol calculations, surface albedo models, photochemical hazes, etc.~will only increase runtimes even more. With this in mind, more recent methods to accelerate retrievals have been explored, such as ultra-efficient radiative transfer schemes \citep{robinson22} and machine learning \citep[e.g.,][]{zingales18, neila18, cobb19, fisher20, HimesEtal2022}. In this study, we achieve an extremely short computational time for Bayesian retrievals using a spectral grid-based method, which was first described in \citet[hereafter S23]{susemiehl23}. In this framework, a grid of model spectra at defined parameter values is pre-generated using a full high-fidelity radiative transfer scheme, and forward models for the Bayesian inference algorithm are then interpolated from the grid. 

In this work, we constrain the detectability of \ce{H2O} absorption in the spectra of Earth-like exoplanets as a function of observational SNR and central bandpass wavelength, in order to lay the groundwork for optimizing exoplanet characterization observations with the Habitable Worlds Observatory. By using the pre-computed grid by S23, we reduce runtime significantly and explore a large span of parameter space (i.e.~SNR by increasing signal and varying molecular abundance) at little computational cost. We present the results of this study and showcase the benefits to grid-based retrieval studies. In $[\S]$ \ref{sec:method} we present the methodology in three parts; a summary of the grid-building work done by S23, a summary of the Planetary Spectrum Generator and the application PSGnest, and a thorough description of our simulations and model parameters. In $[\S]$ \ref{sec:results} we  present the results of our simulations for both the modern Earth-like SNR study and the molecular abundance study. In $[\S]$ \ref{sec:discuss} we discuss the presented results and analyze the impact for future observations of varying Earth-twin epochs. In $[\S]$ \ref{sec:conc} we present our conclusions and ideas for future work.

\section{Methodology}
\label{sec:method}

\subsection{Spectral Grid}
\label{sec:grids}

\begin{figure}[]
\centering
\includegraphics[width=0.51\textwidth]{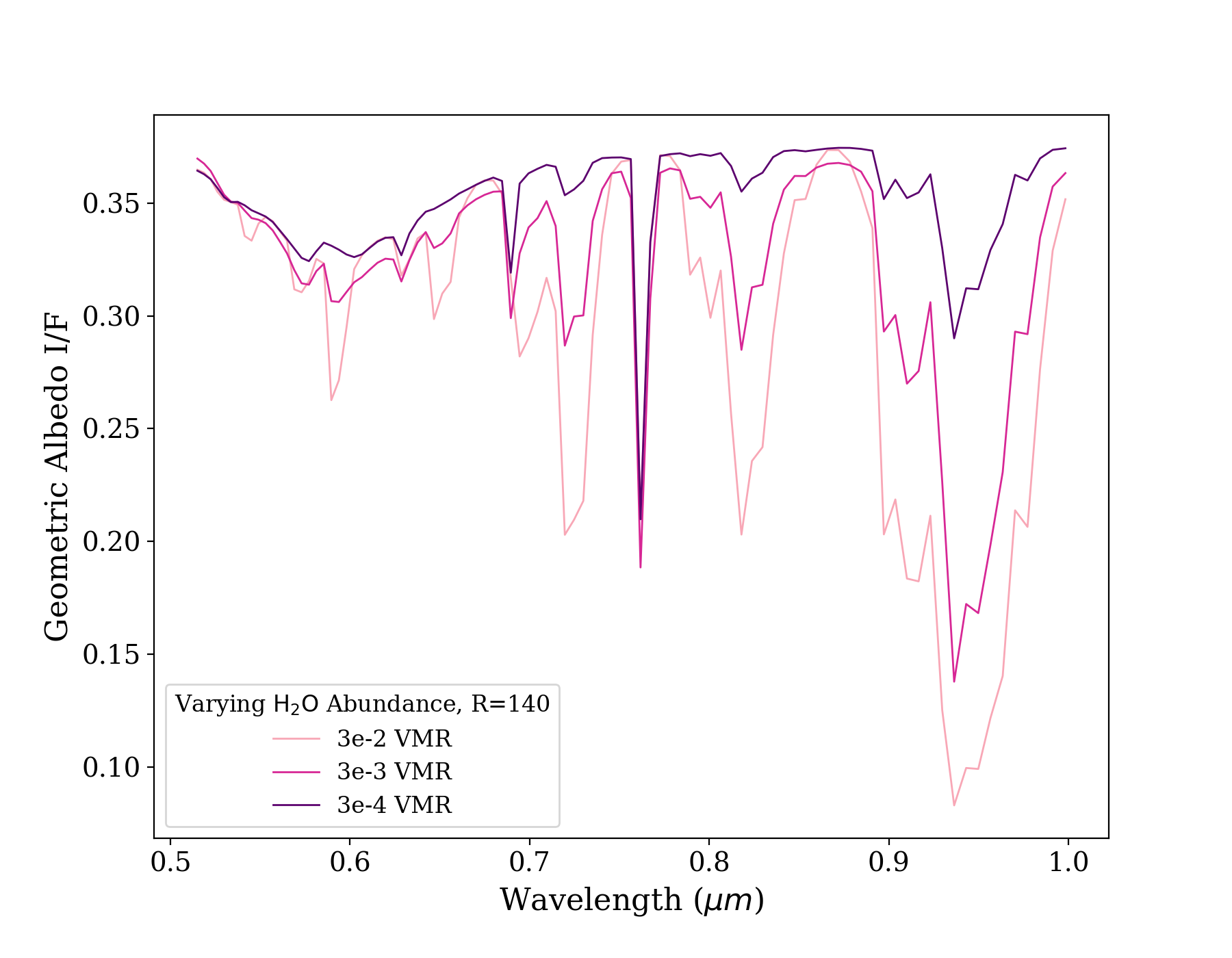}
\caption{Comparison of spectra with varying \ce{H2O} abundance. Shown are the lowest and highest abundance in our simulations (3$\times10^{-4}$ VMR and 3$\times10^{-2}$ VMR) as well as the modern Earth value (3$\times10^{-3}$ VMR). All spectra are binned at R=140. At the highest abundances, \ce{H2O} spectral features begin to overlap.}
\label{fig:spectra_abund}
\end{figure}

 Our ability to perform thousands of nested sampling runs to retrieve exoplanetary atmospheric parameters in a reasonable computational timeframe relies heavily on the efficiency inherent in grid-based retrievals. Using a pre-built grid allows for each retrieval to take on the order of seconds to minutes. 
 
 For this study, we used the grid of rocky planet reflectance spectra built by S23 using the the Planetary Spectrum Generator \citep[PSG,][]{PSG,PSGbook}. PSG is a radiative transfer model and a tool for synthesizing and retrieving planetary spectra, including atmospheres and surfaces for a wide range of planetary properties and covering wavelengths from 50 nm to 100 mm (i.e.~UV/Visible/NIR/IR/FIR/THz/sub-mm/Radio). PSG includes aerosol, atomic, continuum (i.e.~CIAs, Rayleigh, Raman), and molecular scattering and radiative processes; these are implemented with a layer-by-layer radiative transfer framework. For this study and the building of the grid, the latest HITRAN-2020 \citep{gordon20} molecular absorption parameters were used, implemented in PSG with the correlated-k methodology. The HITRAN database also includes collision-induced absorption (CIA) bands, and we characterize the MT\_CKD water continuum as \ce{H2O}-\ce{H2O} and \ce{H2O}-\ce{N2} CIAs \citep{kofmanvilla21}
 
 S23 present a novel approach to grid-building that uses an adaptive grid point approach. The development of the grid featured an algorithm which adds new grid points iteratively at the location of highest interpolation error to reach a specified error contribution based on the expected uncertainty of the data that will be analyzed; the interpolation error is thus reduced at each step by balancing between the peak computational complexity and peak interpolation accuracy. The grids can then be used to produce forward models which are input to the nested sampling routine PSGnest\footnote{https://psg.gsfc.nasa.gov/apps/psgnest.php}. PSGnest is a Bayesian retrieval tool adapted from the Fortran Multinest retrieval algorithm \citep{multinest}, housed in the PSG code base and designed for application to exoplanetary observations. Computation is accelerated via memory-mapping coded in C, enabling ultra-efficient retrievals. The output contains highest-likelihood values of output parameters, the average value resulting from the posterior distribution, uncertainties (also estimated from posterior distribution), and the log evidence (logZ) \citep{PSGbook}.

The current grid of geometric albedo spectra covers a wavelength range from 0.515--1.0 {\microns}, as this range has been defined as the VIS channel in the conceptual exoplanet imaging instruments for future missions such as LUVOIR and HabEx, and was designed to analyze data up to SNR=20 for a spectral sampling of R=140. The grid uses a simpified set of planetary parameters defining the atmospheric composition, with the range of values centered on those of modern Earth as defined in \citet{feng18}. These values consist of constant volume mixing ratios (VMRs) of \ce{H2O}$=3\times10^{-3}$, \ce{O3}$=7\times10^{-7}$, \ce{O2}=$0.21$, a background gas of \ce{N2}=$1-$\ce{H2O}$-$\ce{O3}$-$\ce{O2}, constant temperature profile at 250 K, and surface albedo ($\mathrm{A_{s}}$) of 0.3. For our simulations, the pressure model-top was set to $10^{-4}$ bars with $\mathrm{P_{0}}$ as surface pressure, and the planetary radius was fixed at $\mathrm{R_p}$ = 1 $R_\Earth$. The radius was fixed due to the intrinsic degeneracy between the planetary radius $\mathrm{R_p}$ and a wavelength-independent surface albedo for reflected-light clear sky models; this degenerate effect on the total reflected light of the planet would mean that a specific continuum geometric albedo spectrum could be interpreted as a large planet with low overall albedo or small planet with a high albedo (an issue also addressed in \citealp{feng18}). Specific complementary observations at very short or very long wavelengths or (in rare cases) planetary transit measurements may be able to mitigate this degeneracy and permit the measurement of the absolute radius of the planet; however, for now we have kept the radius fixed in our investigation of on an Earth-like twin, and allowed for a varying surface albedo.

Cloudy versions of each spectrum were also created to complement the clear versions of the same spectrum. The clouds were assumed to be wavelength-independent with a 0.23509 ppm mass mixing ratio and a peak of 1{\microns} for particle size distribution (corresponding to a surface optical depth of 10). We adopt a generic isotropic scattering cloud model appropriate for an Earth-twin observed at a distance, with a cloudiness fraction ($\mathrm{C_f}$) computed by linearly combining a clear ($\mathrm{s_{clear}}$) and cloudy ($\mathrm{s_{cloudy}}$) spectrum: $\mathrm{C_f} \times \mathrm{s_{clear}} + [1-\mathrm{C_f}] \times \mathrm{s_{clear}}$. 

The total grid consists of 1.4 million spectra with parameter points for the parameters laid out in the left column of Table~\ref{tab:params}, along with the minimum and maximum grid values as in Table 1 of S23; the full set of data files for the grid are currently available for use within the PSG nested retrievals application\footnote{https://psg.gsfc.nasa.gov/apps/psgnest.php}. The minimum and maximum grid values are also used as the priors in our simulations. The grid was generated using API calls to a distributed configuration of PSG called GridRunner housed on Goddard Space Flight Center's (GSFC) local Explore cloud computing cluster. In future work, the grid will be rebuilt introducing an increased number of parameters and a wider wavelength range; however, the current pre-built grid allowed for a rapid analysis and is currently available for use on PSG.

\subsection{Simulated Retrievals}
\label{sec:sim}

To explore the optimal bandpass for \ce{H2O} detection as function of SNR, we chose 25 evenly-spaced bandpass centers across the full grid wavelength range of 0.515--1.0 {\microns}. We chose the fiducial parameter set shown in Table~\ref{tab:params} with a bandwidth of 20\% and R$=$140 to be a baseline for comparison. In practice, the bandpass widths are approximately 19\% due to rounding to fully include or exclude the discretized points in the R=140 spectra; however, coronagraphy design for the future HWO observatory is not yet decided and could vary in bandpass width as discussed above. The clear versions of the two spectra are presented in Figure~\ref{fig:spectra_intro} to directly compare the feature smoothing between spectra, along with the resolving power of 70 for reference. One can see that there is loss of smaller features and smoothing of minutia in deeper features, such as the 0.9 {\microns} \ce{H2O} feature as the resolving power decreases. The bandpass centers are presented in the top panel to show the method of visualizing the data, as well as the first and last 20\% bandpasses. It is evident that the bandpasses overlap significantly with each other, resulting in a very thorough study of coronagraphy bandpass possibilities. 
\begin{figure*}[]
\centering
\includegraphics[scale=0.45]{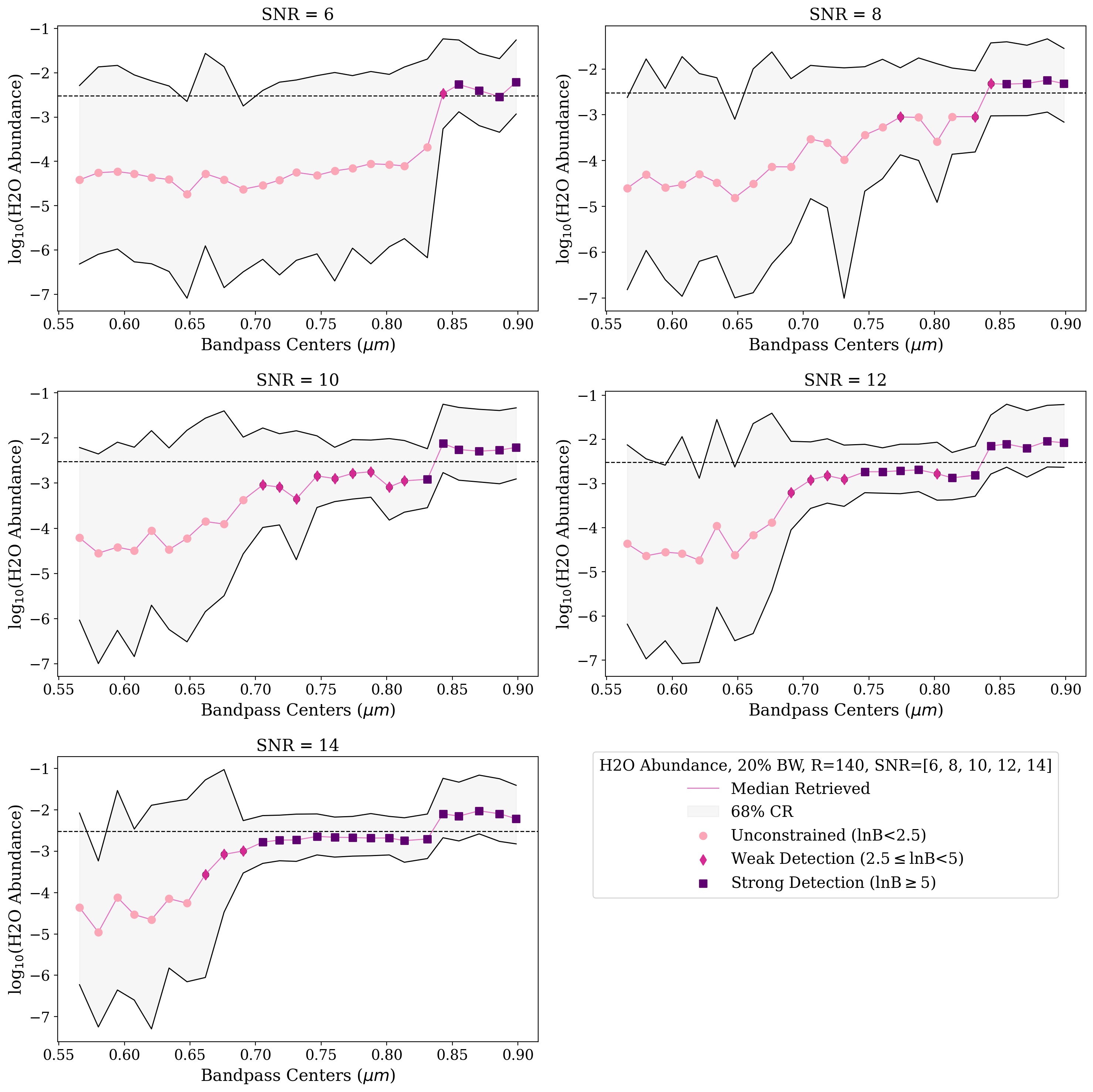}
\caption{Results of the analysis for the fiducial abundance (3$\times10^{-3}$) for \ce{H2O}, as an example case. Each dot represents a bandpass center, with the pink line portraying the median retrieved values and the gray shaded region representing the upper and lower limits for the 68\% credible region; the true value is shown with the black dashed line. Increased statistical certainty can be seen where the gray regions narrow. Each point is colored to indicate varying detection strength. Unconstrained regions ($lnB<2.5$) are shown in light pink dots, weak detections ($2.5 \leq lnB < 5$) are shown in dark pink diamonds, and strong detections ($lnB>5$) are shown in purple squares.}
\label{fig:h2o_bfretest_broad}
\end{figure*} 
There were 400 live points used in PSGnest, and an evidence tolerance of 0.1. For each retrieval performed, the median values as well as the upper and lower limits of the 68\% credible region were calculated by inferring the 1D marginalized posterior probability density function (PDF) using a kernel density estimation (KDE), and then integrating the PDF top-down until summing 68\% of it \citep{harrington22}. We also calculate the Bayes factor for each bandpass retrieval to confirm constraints and detection strength for each retrieved atmospheric constituent. The Bayes factor is calculated by subtracting the Bayesian log-evidence per retrieval using the previously-assumed gas abundances \citep{trotta08}; the differences in log-evidences is the log-Bayes Factor, $\mathrm{lnB}$ \citep{benneke13}. If $\mathrm{lnB}$ is less than 2.5, it represents an unconstrained detection for that data simulation; if $\mathrm{lnB}$ is between 2.5 and 5 a weak detection is represented, and if $\mathrm{lnB}$ is greater than 5 a strong detection is represented \citep[reference Table 2 of][]{benneke13}. For our simulations, the comparison occurs between a retrieval with all parameters included and retrieved versus a retrieval with one abundance parameter (the parameter of interest) removed from the retrieval and the abundance set to zero. The 68\% credible region contains the true value 68\% of the time, whether or not the molecule is present, whereas $\mathrm{lnB}$ directly examines if an improved fit occurs with the molecule included. Thus, we prioritize $\mathrm{lnB}$ as it directly addresses the likelihood of the presence of the molecular species and its detectability. The log-Bayes Factor is not calculated for non-gaseous parameters as the values are not associated with a presence or absence of evidence.
\begin{figure*}[]
\centering
\includegraphics[scale=0.45]{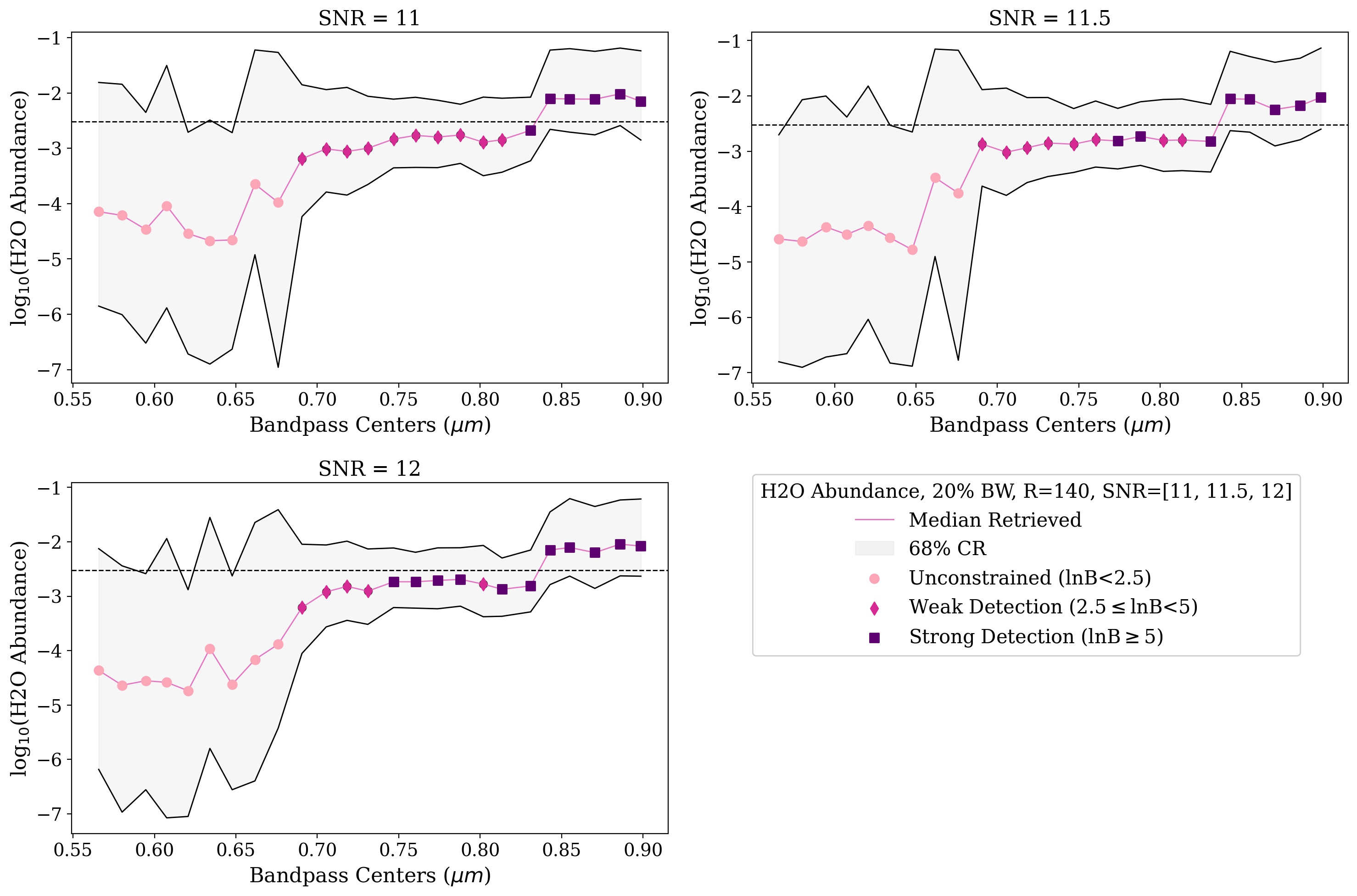}
\caption{Results of the analysis for the fiducial abundance for \ce{H2O}, but for a narrower SNR range. All facets of the plot remain the same as in Figure~\ref{fig:h2o_bfretest_broad}. Detectability at wavelengths below 0.84{\microns} occurs for SNR$>$11.}
\label{fig:h2o_bfretest_narrow}
\end{figure*}
We also examined a range of assumed signal-to-noise ratios (SNRs) for the data set. Since exoplanets are so much fainter than their host stars (i.e.~10 orders of magnitude for the Earth-Sun system), observations are going to be extremely challenging and likely require long observing times. Signal increases linearly with time, whereas noise increases with the square root of the time (in the source-noise-dominated regime). The combined atmospheric signal across all wavelengths will have to be much stronger than the standard deviation of the noise in order be able to derive any meaningful constraints on the planet's atmosphere. Thus, we delve deeply into the study of SNR and the derivable constraints and detections as signal increases. The noise term for the SNR here refers to the standard deviation of the noise at all pixels and is added as Gaussian noise after the simulations of the spectra to simply mimic SNR in real observations. Similar to \citet{feng18}, who studied an SNR range of 5, 10, 15, 20, we studied a preliminary broad range (3 - 16 in steps of 1), and then added smaller SNR steps (8.5, 9.5, 10.5, 11.5). We explore a broader range of SNR to thoroughly investigate the change in \ce{H2O} detectability and the associated detection strength, specifically between SNRs of 10 and 12. 

To add another dimension to our study, we explored the effect of varying molecular abundances on detectability as a function of SNR. Figure~\ref{fig:spectra_abund} portrays how varying the abundance of \ce{H2O} influences the spectrum. One can see that as the abundance of \ce{H2O} increases, the feature depth also increases, as expected. The lowest abundance of 3$\times10^{-4}$ VMR has very shallow features, with increasing depth moving up to 3$\times10^{-2}$ VMR, the highest abundance. To finely explore the influence of each molecule, we only vary the abundance of the molecular species of interest and maintain the other components at the modern Earth-like values as presented in Table~\ref{tab:params}. We vary our molecular abundances by stepping through log-space, increasing or decreasing by 0.25 and, at larger values, 0.5. We center around the modern Earth-like values for each molecule, to provide a familiar starting point. This also means that although the physical chemistry of each model is not accurate to any specific epoch of Earth, we can match the individual abundances as closely as possible to various epochs of interest. Our abundance values range from 3$\times10^{-4}$ to 3$\times10^{-2}$ VMR. 

For the purposes of this study, we will be focusing on our results for the \ce{H2O} abundance study, with a mention of the preliminary results found for \ce{O2}. We note that the absolute SNR results are subject to change with newer grid construction, due to upgrades in PSG and the addition of more parameters and molecules; the results of this study are therefore most valuable for understanding the relationships between bandpass wavelength, SNR and \ce{H2O} abundance with regard to detectability. Future work will explore the impact of more physically realistic atmospheric compositions and the retrieval of additional atmospheric parameters, and we will produce updated spectral grids accordingly.

Each case study (i.e.~every SNR in the previous range applied to each abundance study) took approximately 5 minutes to run on the GridRunner application of PSG; comparatively, each case study on a local machine (MacBook Air 2017, 1.8GHz dual-core Intel Core i5) took approximately 1.5 to 2 hours. Each case study consists of thousands of spectra to be retrieved upon, thus GridRunner massively accelerated our progress and saved days of time. 

\section{Results}
\label{sec:results}

\subsection{Modern Earth Case Results}
\label{sec:modern}

We begin by presenting the detectability of \ce{H2O} as a function of SNR for the fiducial modern Earth case, as first examined in S23. We present a portion of the SNR range in Figure~\ref{fig:h2o_bfretest_broad}; all of the \ce{H2O} data and calculated log-Bayes Factor across abundance, SNR, and wavelength are available to the community on Zenodo\footnote{10.5281/zenodo.7897198}. As SNR increases, the detectability of smaller \ce{H2O} spectral features at shorter wavelengths (such as the 0.74 {\microns} feature) improves; this is evident in the narrowing of the 68\% credible region for the strong detection points in Figure~\ref{fig:h2o_bfretest_broad}. There is a significant change in detectability at shorter wavelengths between an SNR of 8 and an SNR of 10 --- at SNR = 8 we achieve a strong detection at 0.9 {\microns} but it is unconstrained at all other points, while for SNR = 10 we find weak detections down to 0.71 {\microns}. There is also a tipping point of diminishing returns at approximately SNR = 14, after which there is no change in the bandpasses where \ce{H2O} is detectable and very slight changes in the sizes of the credible regions. There is a notable shift in detectability between SNR = 11 and SNR = 12, wherein the 0.74 {\microns} water feature goes from a firmly weak detection to a firmly strong detection with a narrower credible region. Thus we investigated further with smaller SNR steps in this transition region to fully understand the shift in detectability, portrayed in Figure~\ref{fig:h2o_bfretest_narrow}; for example, at SNR = 11.5 the 0.74 {\microns} bandpass center begins to yield a strong detection of \ce{H2O} where previously it was a weak detection. 

\begin{figure}[]
\centering
\includegraphics[width=0.5\textwidth]{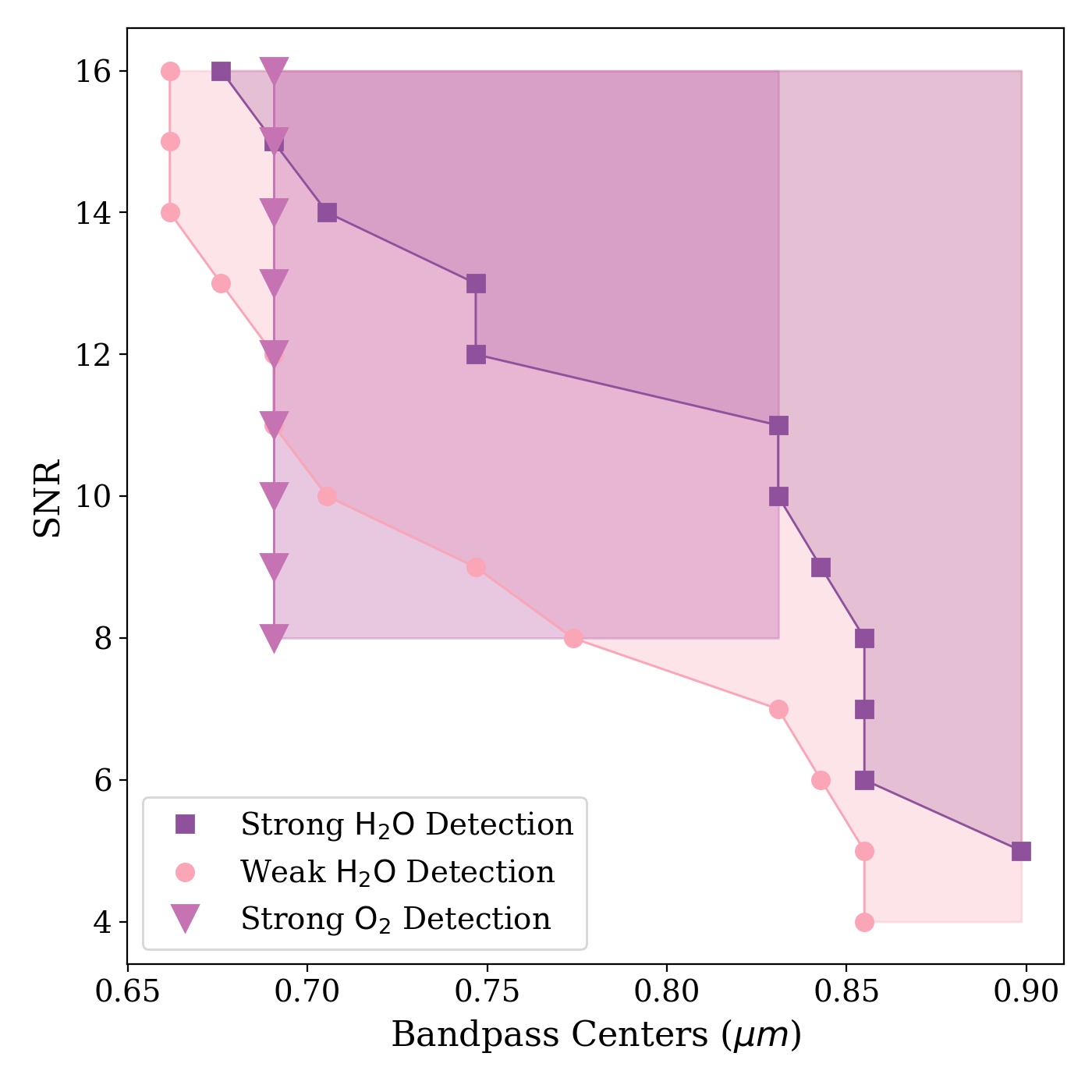}
\caption{Summary of the bandpasses where \ce{H2O} and \ce{O2} are detectable for our fiducial Modern Earth scenario, based on results as shown in Figures~\ref{fig:h2o_bfretest_broad} and \ref{fig:h2o_bfretest_narrow}. The shortest bandpass center at which one can achieve a strong or weak detection for \ce{H2O}, or a strong detection for \ce{O2}, are shown in dark purple squares, pink dots, and light purple triangles, respectively. SNR is on the y-axis, and the bandpass centers are on the x-axis. We also show the range out to the longest wavelength at which the same detection is achieved as shaded regions. We note that a strong detection of both \ce{H2O} and \ce{O2} can be achieved with bandpasses at or below 0.83 {\microns} and SNR$\ge$10.}
\label{fig:shortest_detec_h2o}
\end{figure}
As we discussed in \S\ref{sec:intro} and is demonstrated in Figures~\ref{fig:h2o_bfretest_broad} and \ref{fig:h2o_bfretest_narrow}, the minimum wavelength for detection is directly related to the SNR of the data, but longer wavelengths also yield detections due to the increasing depth of the \ce{H2O} spectral features. We summarize this result in Figure~\ref{fig:shortest_detec_h2o}, where we plot the positions of the minimum bandpass where either a strong or weak detection of \ce{H2O} can be achieved for a specified SNR; the results show an almost linear relationship between the shortest detectable bandpass center (at either weak or strong detection) and the SNR value required. We note that there are no \ce{H2O} detections for wavelengths shorter than 0.66 {\microns} since there are no water vapor features available shortwards of this bandpass, and for SNR $>$ 14 we are strongly detecting \ce{H2O} at all wavelengths longer than 0.65 {\microns}. To show the full range of detectability, we shade the region out to the longest wavelength at which a detection is achieved. \ce{H2O} is always strongly detectable at the 0.9 {\microns} feature, leading to larger likelihood of \ce{H2O} detectability as we move to higher SNRs. The weak detection of \ce{H2O} shows a similar result, with 0.9 {\microns} weakly detectable to the lowest SNR in our study. For comparison, we also show a similar analysis for \ce{O2}, which only has one strong spectral feature at 0.76 {\microns} in our spectral range; therefore above SNR = 8, the longest and shortest wavelengths at which \ce{O2} is detectable are constant. 

\begin{figure*}
\gridline{\fig{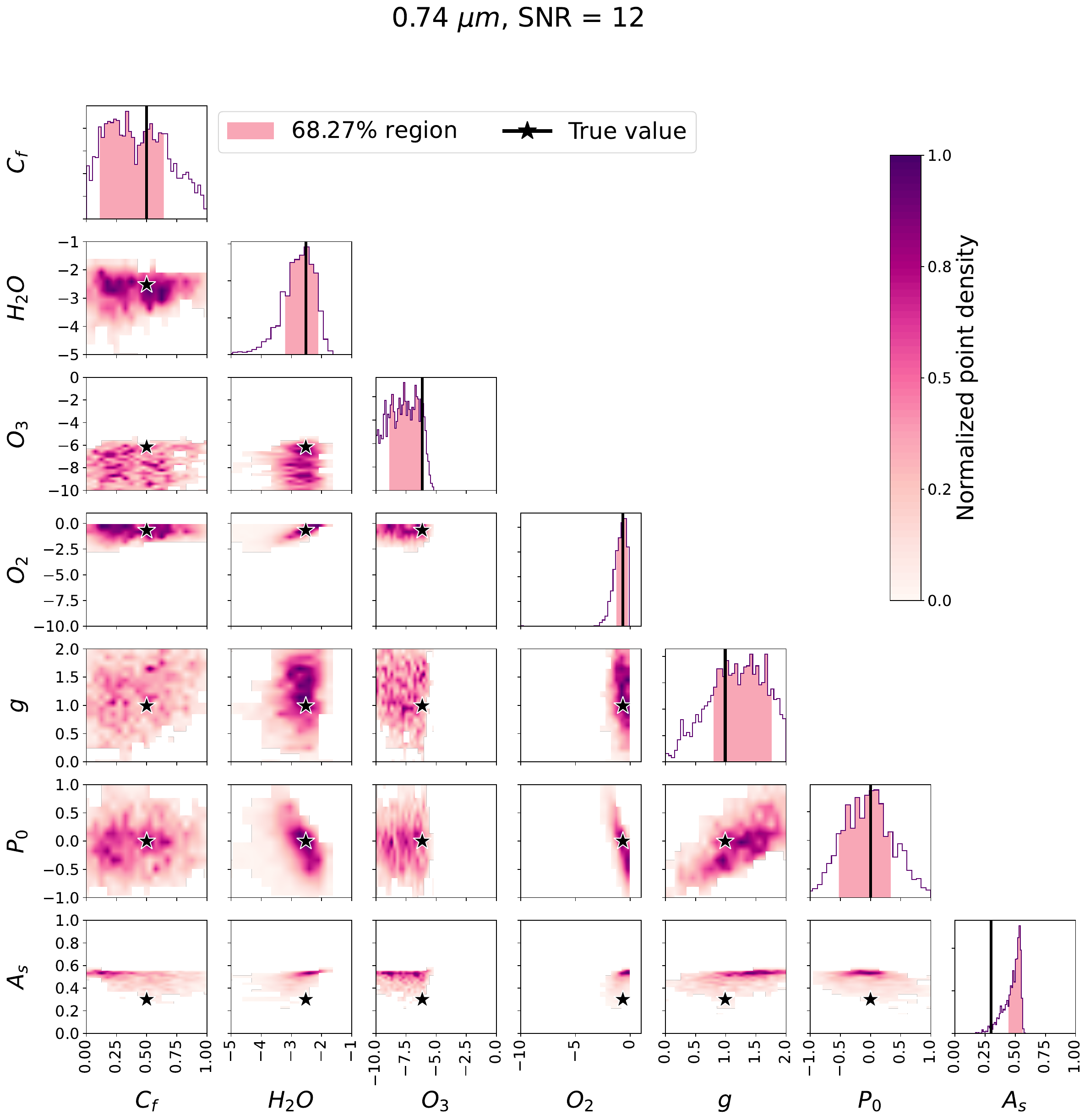}{0.45\textwidth}{(a)      
          Corner plot for the bandpass centered on 0.74 {\microns}.}
          \fig{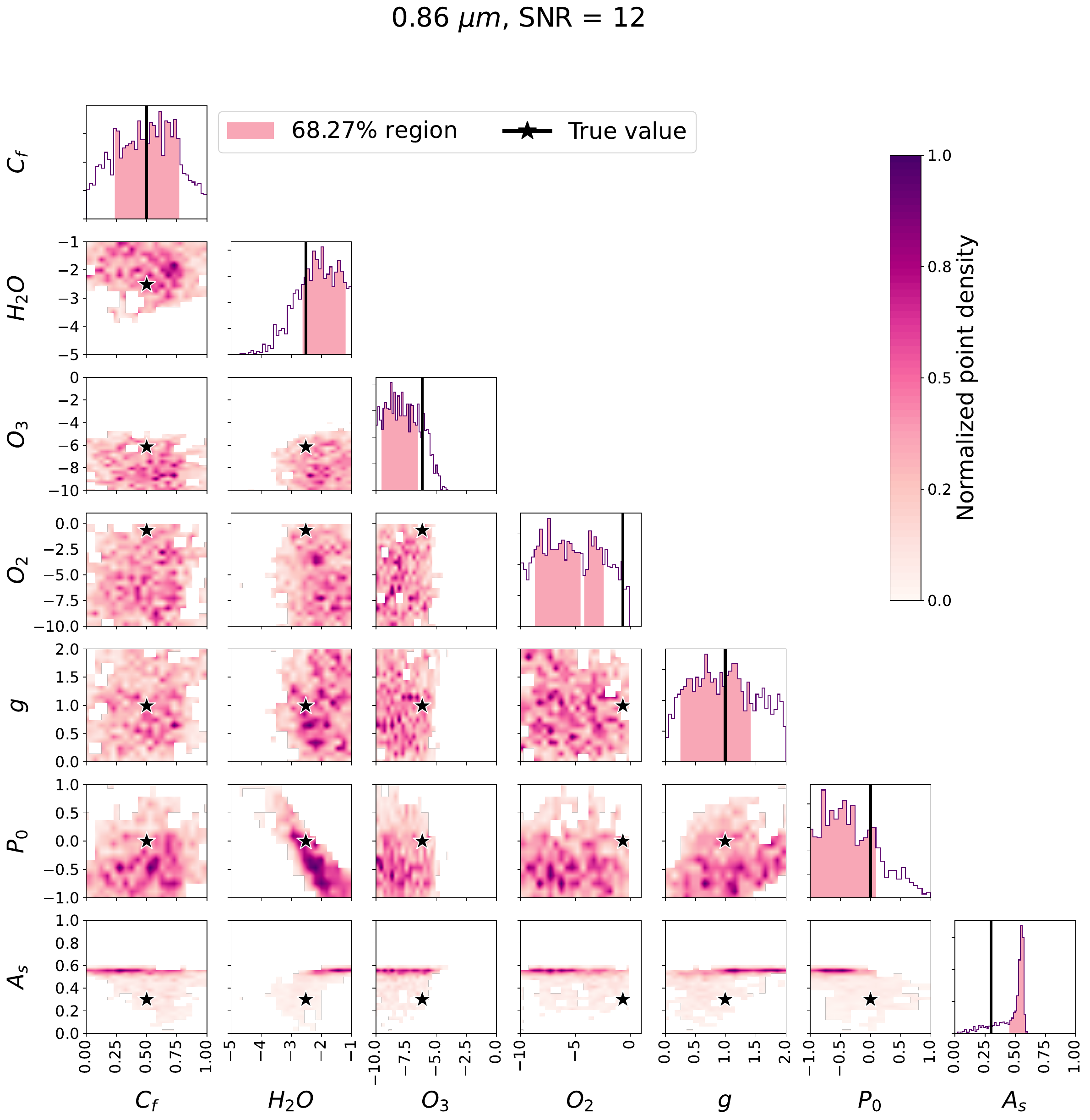}{0.45\textwidth}{(b) Corner plot for the bandpass centered on 0.86 {\microns}.}}
\gridline{\fig{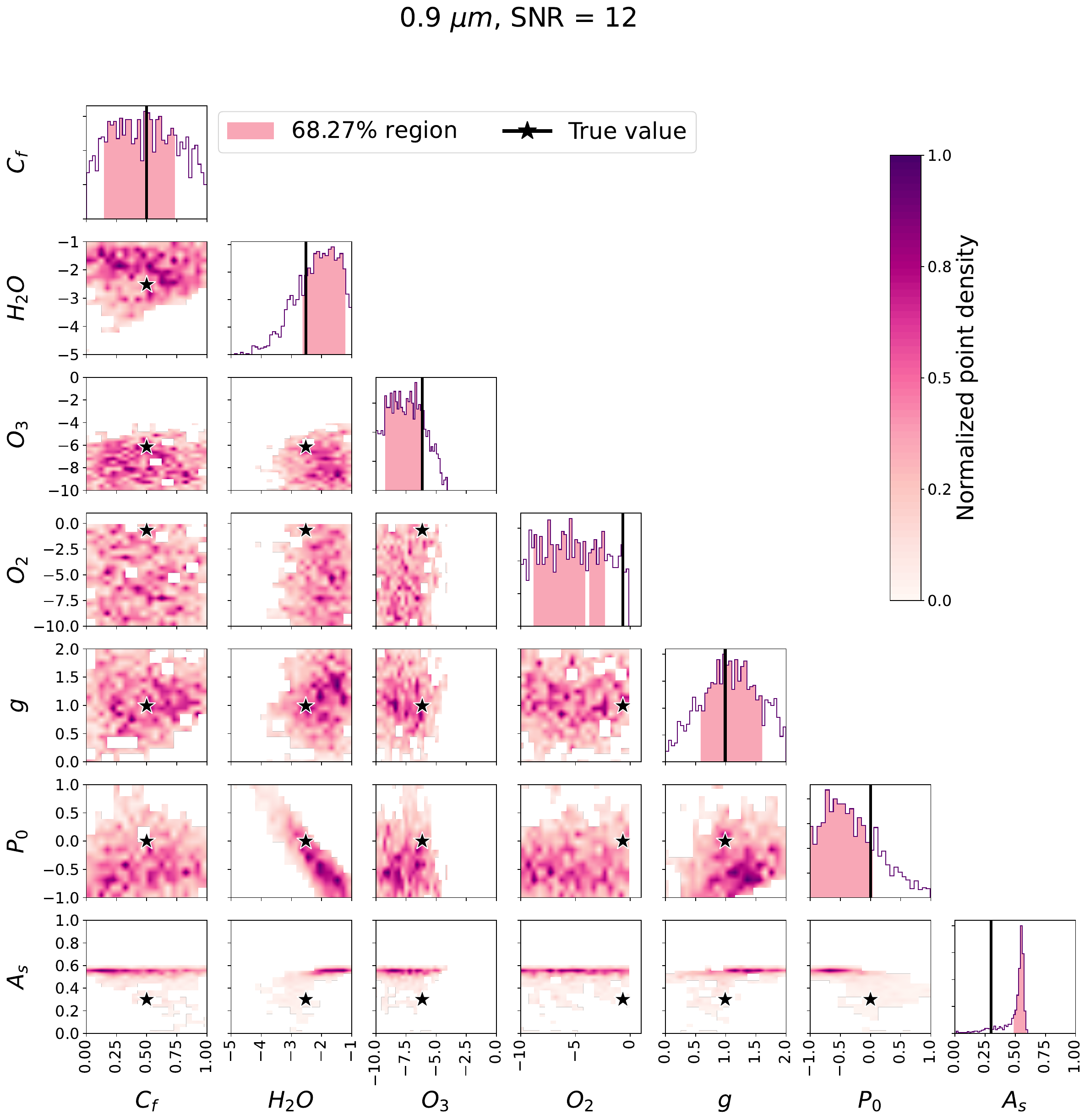}{0.45\textwidth}{(c)     
          Corner plot for the bandpass centered on 0.9 {\microns}.}
          \fig{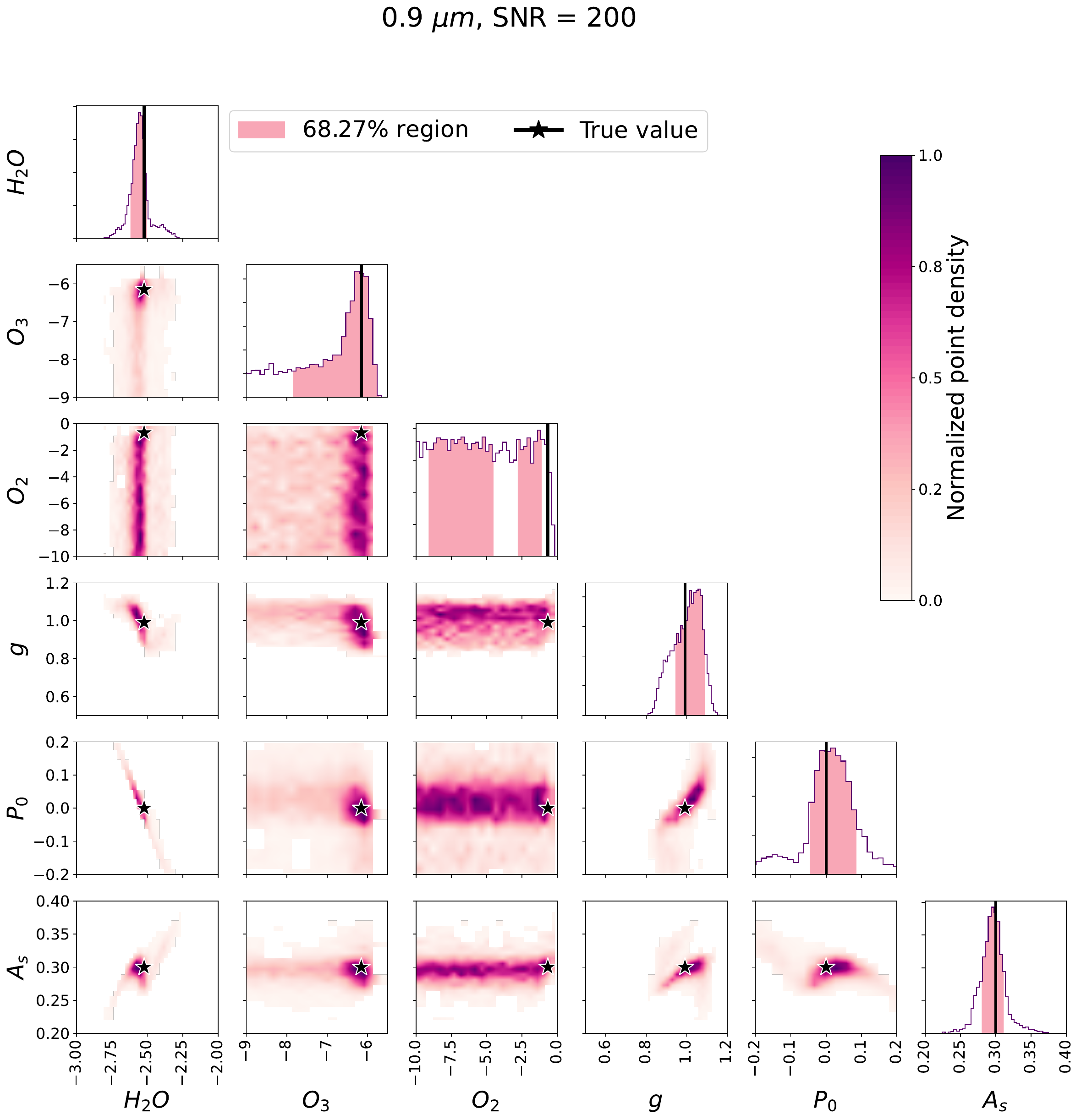}{0.45\textwidth}{(d) 0.9 {\microns} corner plot with SNR of 200, locked $\mathrm{C_f}$.}}
\caption{Three corner plots, one for each notable \ce{H2O} feature, with one corner plot at very high SNR to portray the intrinsic parameter degeneracies. Each has the same parameters with SNR of 12, shown in Table~\ref{tab:params}, with the exception of locking $\mathrm{C_f}$ and an SNR of 200 for corner plot d. In all corner plots, the 68\% credible regions are shown as pink shading in the 1D marginalized posterior distributions along the diagonal of the corner plot, and the true values are represented by black lines in the diagonals of the corner plot, and black stars within the 2D plots.} \label{fig:corner}
\end{figure*}

We note that as we approach the longest wavelengths in the visible range, we notice that the true value of \ce{H2O} is not retrieved, but rather finds a higher abundance of \ce{H2O} than the true value. This is due to a degeneracy with pressure in the last three bandpasses driven by a lack of continuum due to the depth of the water features; in other words, the water absorption region is saturated and wide enough that the continuum cannot be correctly retrieved. This is confirmed by investigating the corner plots made for bandpasses covering different \ce{H2O} spectral features. 

In Figure~\ref{fig:corner}a, which shows the corner plot for the retrieval results for the 0.74 {\microns} bandpass, the high-probability region is perfectly aligned with the true values for $\mathrm{P_{0}}$ and \ce{H2O}; in contrast, we see in Figure~\ref{fig:corner}b that the 0.86 {\microns} bandpass fares slightly less well compared to the 0.74 {\microns} bandpass, with the high-probability regions not as well constrained, but still aligned in part to the true values. However, the high-probability region here is much more elongated, still leading to uncertainty and inaccurately retrieved values, with the true values found at the edges of the 1D 68\% credible regions. Comparing these figures to the corner plot for the 0.9 {\microns} bandpass in Figure~\ref{fig:corner}c, the difference is noticeable. The high-probability region shown between surface pressure $\mathrm{P_{0}}$ and \ce{H2O} is distanced away from the true value, which is found to be on the tail of the high probability region and again at the edges of the 1D 68\% credible regions; this leads to a poorer constraint on \ce{H2O} in selected bandpasses with longer wavelengths, despite the fact that they cover the largest water feature in the visible wavelength range, with a much better constraint at shorter wavelengths. Interestingly, the lack of continuum and related degeneracy does not affect the ability to detect \ce{H2O} at any wavelength, nor does it affect the strength of the detection, as we can see from Figures~\ref{fig:h2o_bfretest_broad} and \ref{fig:h2o_bfretest_narrow}. It is also important to note that there is also a degeneracy with albedo ($\mathrm{A_{s}}$) throughout the full spectrum, which leads to a three-way degeneracy at longer wavelengths. 

Another way to investigate this degeneracy is to effectively remove the impact of spectral noise in order to investigate the impact of parameter degeneracies alone. We increased the SNR to 200, thus minimizing posterior broadening due to SNR, and locked $\mathrm{C_f}$ to examine whether that is a factor in driving \ce{H2O} from the true value. We present the corresponding corner plot in Figure~\ref{fig:corner}d (note that the axis ranges have changed in order to visualize the degeneracies). Without the noise introduced by low SNR, we can vividly see the degenerate relationships between variables, such as $\mathrm{P_{0}}$ and \ce{H2O}; although both are well within 1-$\sigma$ of the true values, there are still elongated probability regions. \ce{O2} is poorly constrained, as expected due to the absence of the molecule in this region, while gravity becomes very well constrained at high SNR without $\mathrm{C_f}$ interference. The impact of variations in cloud coverage on \ce{H2O} detectability was not investigated in detail in this work, but earlier work such as \citet{kawashima19} has shown that clouds can increase the detectability of atmospheric species due to the increase in overall planet brightness; however, for \ce{H2O} this is only relevant for extreme cases of high cloud coverage at high altitudes, as the 0.9 {\microns} \ce{H2O} feature has significant depth regardless of cloud coverage.

\begin{figure}
\gridline{\fig{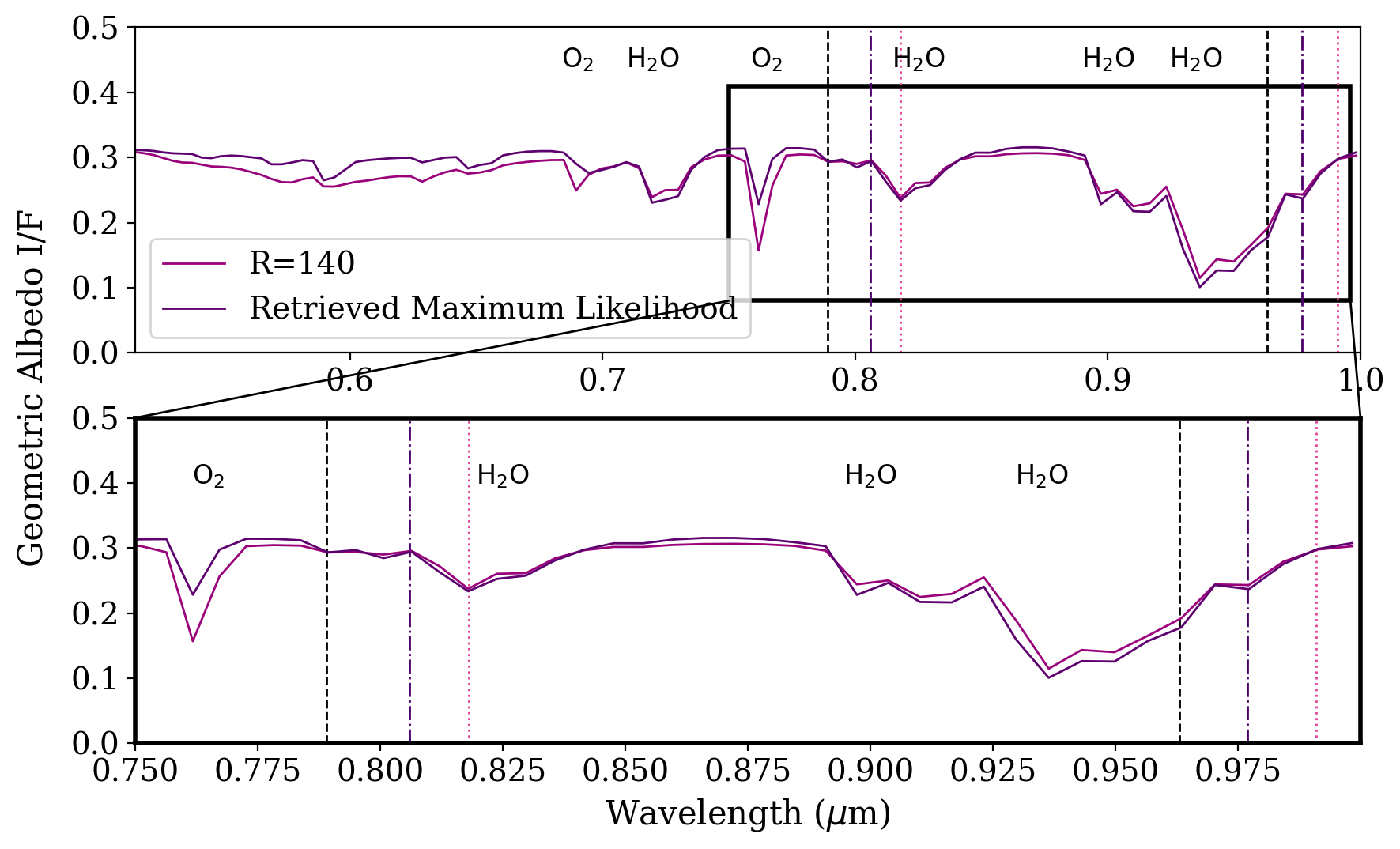}{0.49\textwidth}{(a)         
          This plot zooms in on the 0.9 {\microns} selected bandpasses.}}
\gridline{\fig{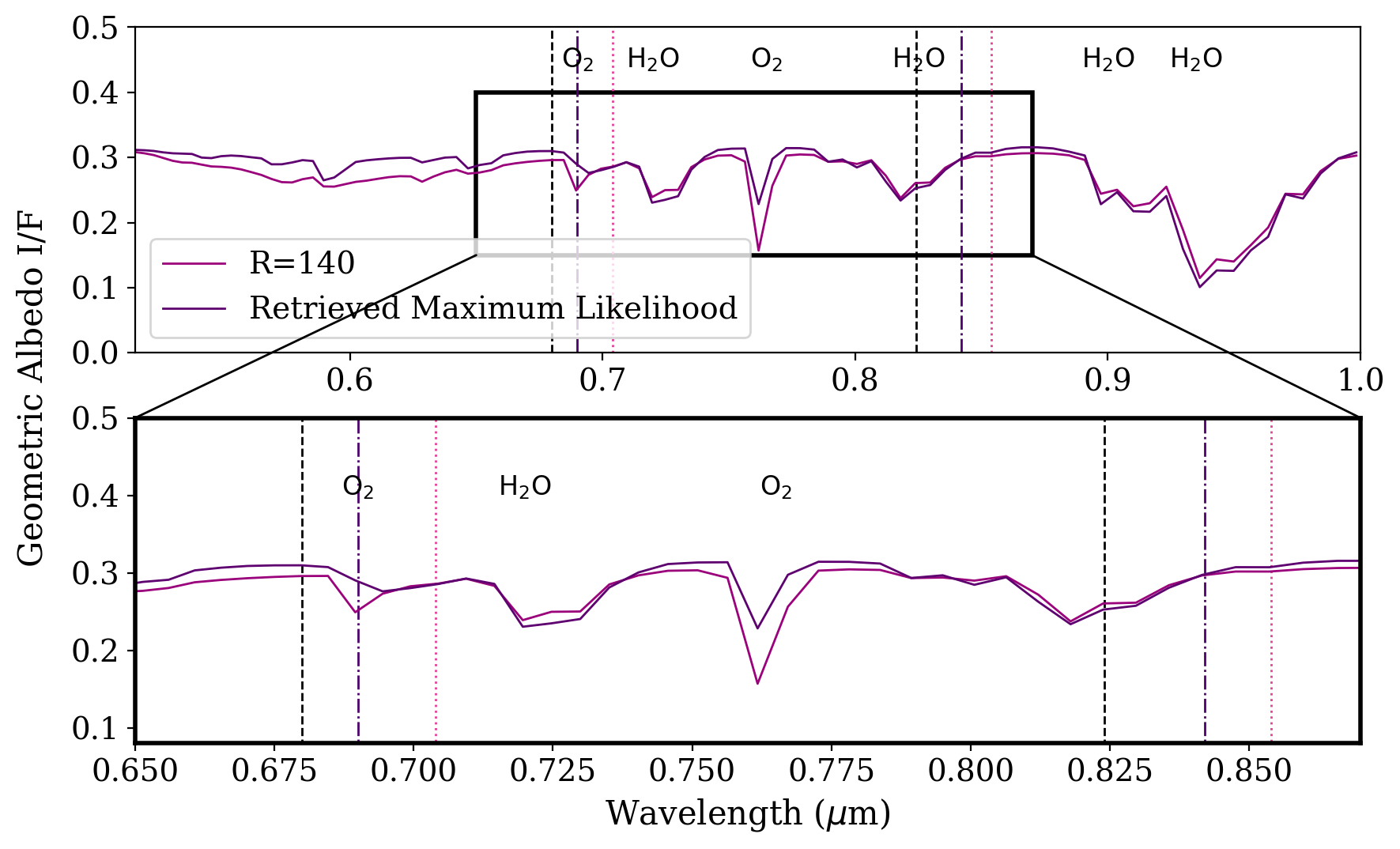}{0.49\textwidth}{(b) This plot zooms in on the 0.74 {\microns} selected bandpasses.}}
\caption{Default spectrum used for simulations, plotted with the high-probability values as drawn from the corner plots and posterior values of $\mathrm{P_{0}}$ and \ce{H2O}. The top panel shows a zoomed-in view of the longer wavelength ranges, and the bottom panel is identical but zooms in to the shorter wavelengths. Each portray the three bandpasses that are most centered with the features of interest. The presence of \ce{O2} in the short-wavelength region helps to break the degeneracies between multiple parameters. \label{fig:spectra}}
\end{figure}

To further illustrate this degeneracy, Figures~\ref{fig:spectra}a and \ref{fig:spectra}b show the full-wavelength spectrum with all the input modern-Earth values, plotted with a spectrum with $\mathrm{P_{0}}$ and \ce{H2O} replaced with the maximum probability values shown in Figure~\ref{fig:corner}c (approximately -0.4 for $\mathrm{P_{0}}$ and -1.6 for \ce{H2O} in log-space). Each figure also presents the three bandpasses most central to the wavelength of interest. One can clearly see that the maximum probability spectrum mirrors the spectrum in the 0.9 {\microns} section almost perfectly, which is what one would expect following the corner plot. However in the 0.74 {\microns} region the spectrum fits poorly, specifically in the regions with \ce{O2}. This contribution of \ce{O2} allows the code to differentiate and break the $\mathrm{P_{0}}$ and $\mathrm{A_{s}}$ degeneracies. Thus we see the results presented previously, wherein abundances are well constrained in shorter wavelengths (with more molecular species present) in contrast to longer wavelengths (where there is only water present). This occurs at all SNR, however the credible region changes as SNR increases, and we see that the true value of \ce{H2O} shifts to the edge of the credible region for longer wavelengths, most notably at an SNR of 14, and at higher SNRs is shown to still be retrieved at the edge of the 1D 68\% credible region in Figure~\ref{fig:corner}d. As such, although we can strongly detect \ce{H2O} at longer wavelengths with low SNRs, the trade-off is accuracy in abundance measurement. However, the insufficient credible region could also be affected by the known deficiencies in Multinest \citep{buchner14, ardevol22, himes22}; we will explore this possibility in more depth in future work.

\subsection{Results for Varying Abundance Cases}
\label{sec:abun}

\begin{figure}
\gridline{\fig{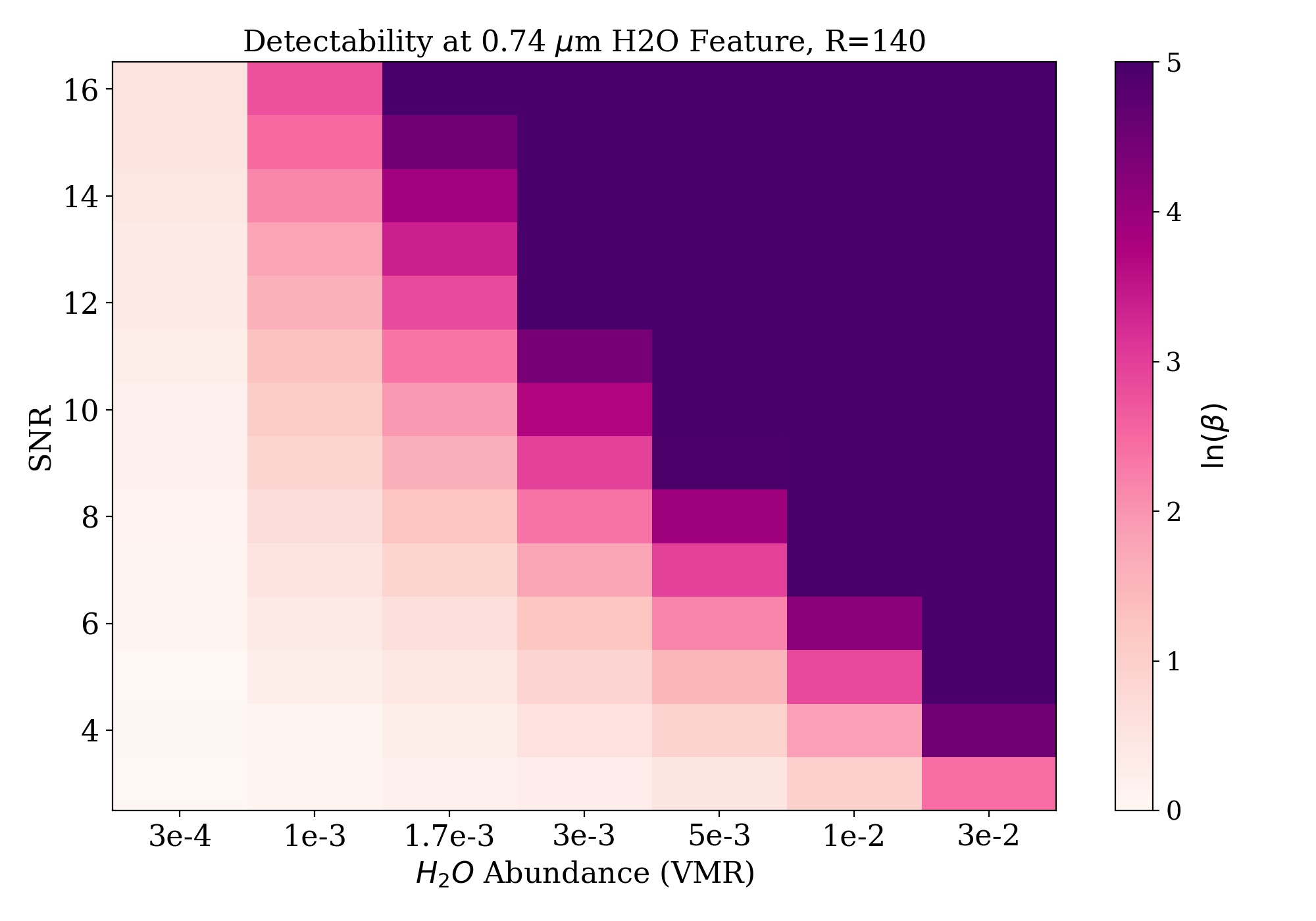}{0.49\textwidth}{(a)         
          Detectability for the 0.74 {\microns} water feature at all abundances in the case study.}}
\gridline{\fig{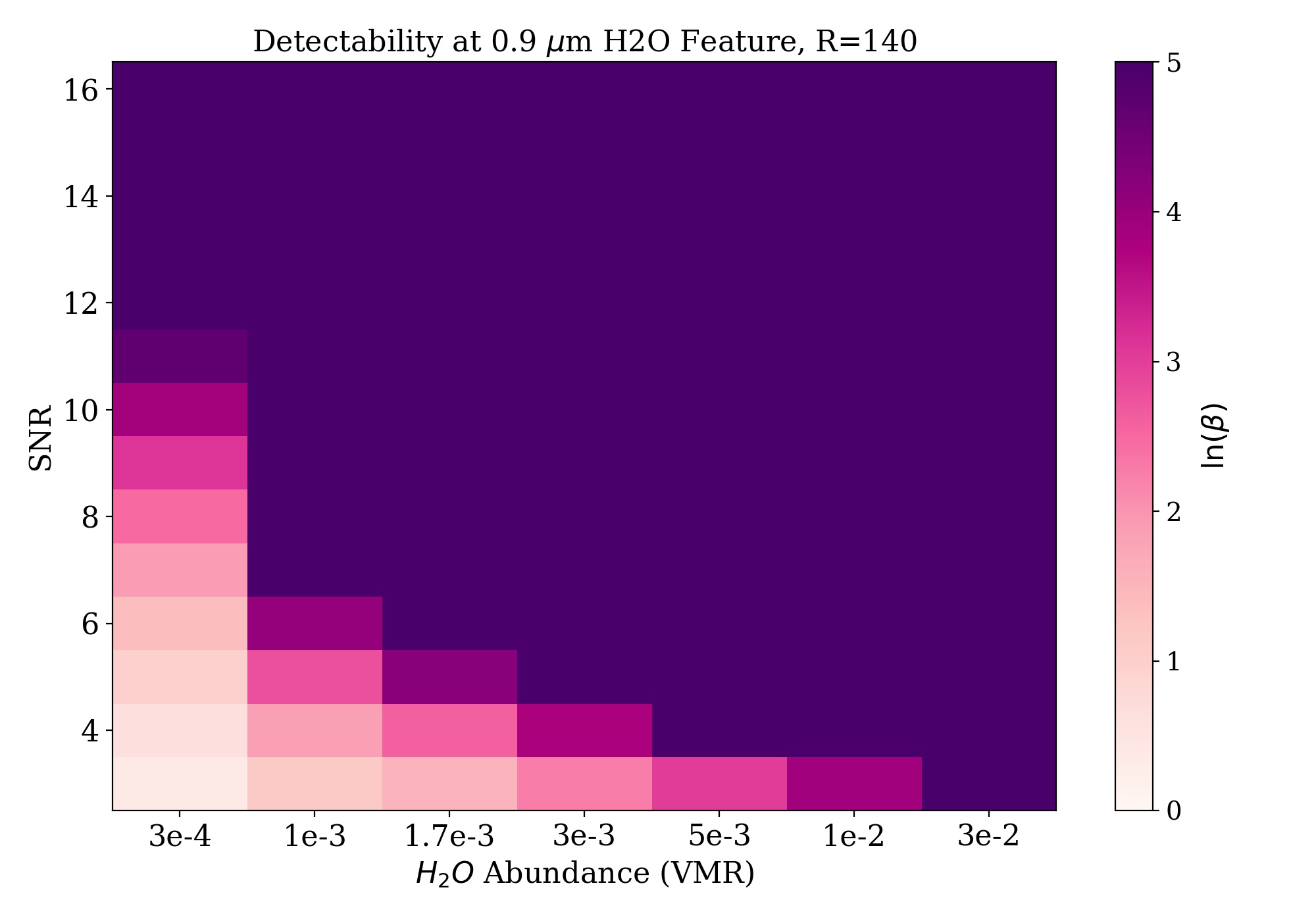}{0.49\textwidth}{(b) Detectability for the 0.9 {\microns} water feature at all abundances in the case study.}}
\caption{Heatmap plots illustrating detection strength as a function of SNR and varying \ce{H2O} abundance. SNR is on the y-axis, \ce{H2O} abundance is on the x-axis, and the colorbar shows the range of log-Bayes Factor ($\mathrm{lnB}$) from 0 to 5 to describe detection strength. $lnB<2.5$ are unconstrained, $2.5 \leq lnB < 5$ are weak, and $lnB>5$ are strong, as in prior figures. \label{fig:heatmaps_h2o}}
\end{figure}

At this point in our study, we shift to present our abundance case study, wherein we vary the abundance of \ce{H2O} above and below modern-Earth values. To assess the trade-off between longer observations (i.e.~stronger signals and lower noise) and different concentrations of \ce{H2O}, we varied the SNR on the observations for the full range of different \ce{H2O} VMRs. As examples, in Figures~\ref{fig:heatmaps_h2o}a and ~\ref{fig:heatmaps_h2o}b we present two heat maps summarizing the \ce{H2O} detectability for the 0.74 {\microns} bandpass and the 0.9 {\microns} bandpass. Both color bars cover the full range of results between 0 and 5 log-Bayes factor, to represent the change in detection strength from weak to strong. Focusing first on Figure~~\ref{fig:heatmaps_h2o}b, we can see that for almost all abundances of \ce{H2O} it is possible to achieve a strong detection. At a VMR of 3$\times10^{-4}$, it requires an SNR = 12 to achieve a strong detection. There is a significant improvement beyond VMR = 1$\times10^{-3}$, thus resulting in a strong detection at SNR of 7. Thereafter, the SNR required for strong detection decreases linearly as abundance rises. We can see that as a general pattern, detection using the 0.9 {\microns} bandpass is achievable for a wide range of abundances at relatively low SNR values. Looking now to Figure~\ref{fig:heatmaps_h2o}a, we see a far different story --- strong detection for the 0.74 {\microns} bandpass does not occur until a VMR of 1.7$\times10^{-3}$, and that occurs at SNR = 16. It is at this point that we see the same linear progression as in the top panel, with decreasing SNR and increasing abundance, until we once again study the highest abundance tested in the study (3$\times10^{-2}$ VMR) and we see a strong detection occur at SNR of 5. 

\begin{figure}[]
\centering
\includegraphics[width=0.5\textwidth]{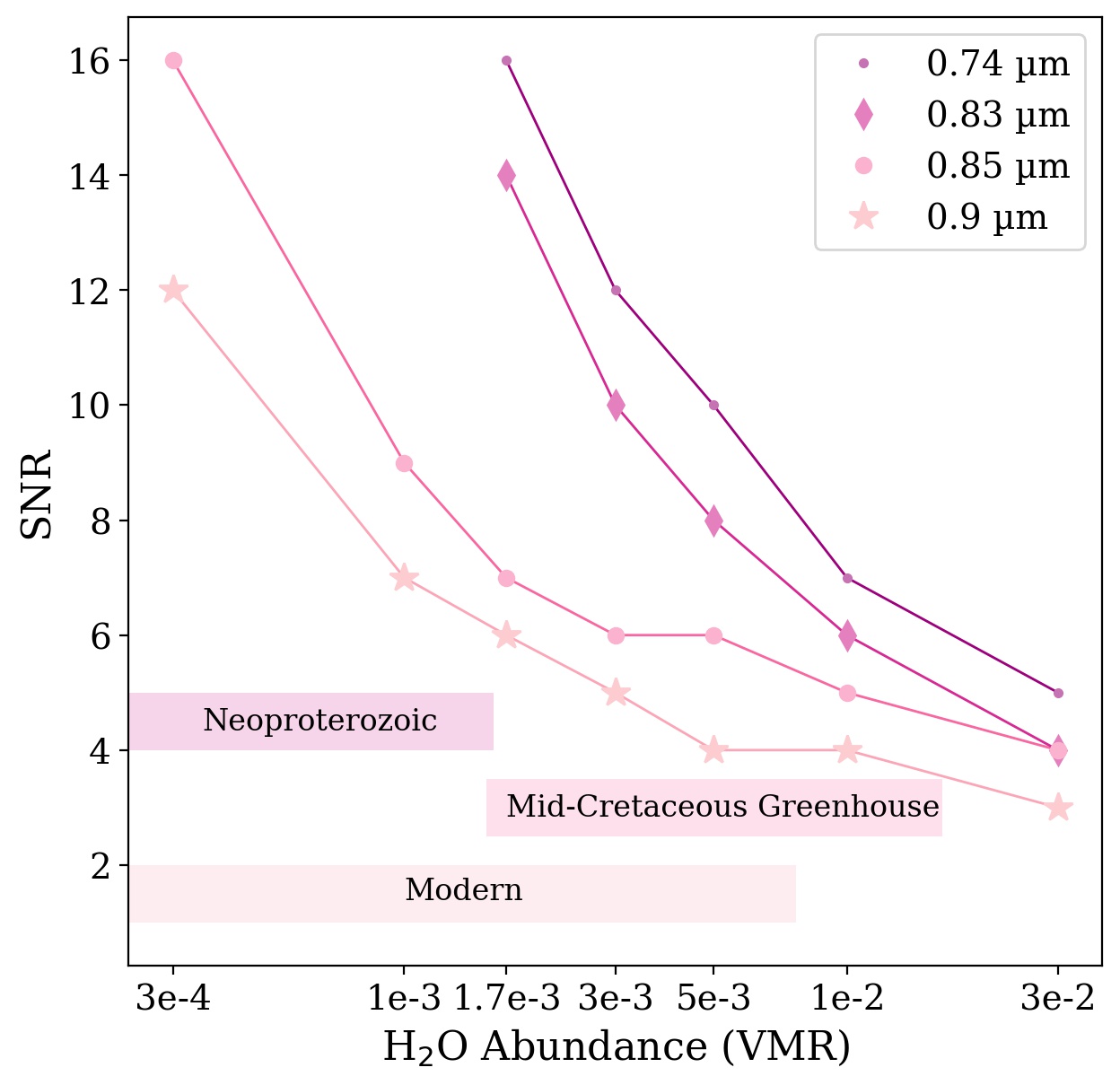}
\caption{Plot of the lowest SNR values at which a strong detection is achieved for notable \ce{H2O} feature bandpasses, as a function of \ce{H2O} abundance. The VMR values are on the x-axis, with SNR on the y-axis. The 0.9, 0.83, 0.85, and 0.74 {\microns} bandpasses are shown in light pink stars, pink dots, dark pink diamonds, and purple dots, respectively. We also mark the rough \ce{H2O} abundance ranges for several epochs through Earth's history, including a Neoproterozoic Slushball period (in dark pink), a Mid-Cretaceous Greenhouse warm period (pink) and the measured range for Modern Earth values (light pink). We note that for display purposes, we have left off the lowest abundance values for the Neoproterozoic Slushball and Modern epochs from the plot range.}
\label{fig:strong_quad_epochs}
\end{figure}

In Figure~\ref{fig:strong_quad_epochs} we display the minimum SNR required to achieve a strong detection for each abundance of \ce{H2O} for four representative bandpasses across our wavelength range - 0.9, 0.85, 0.83, and 0.74 {\microns}. We notice that at very low abundances of \ce{H2O}, it requires a high SNR even for the 0.9 {\microns} bandpass for a strong detection, and it is not strongly detectable at any wavelengths lower than 0.85 {\microns}. Conversely, at very high abundances of \ce{H2O}, every selected bandpass with \ce{H2O} is strongly detectable at low SNRs, with a highest required SNR of 5 for \ce{H2O} at 0.74 {\microns}. To provide context to the results, we present ranges for the zonally averaged abundance of \ce{H2O} for several epochs of Earth's history derived from global circulation models (GCMs), including a lower-CO$_2$ Neoproterozoic Slushball scenario (minimum VMR of 3.1$\times10^{-5}$, maximum VMR of 1.6$\times10^{-3}$), a warmer Mid-Cretaceous Greenhouse period (minimum VMR of 1.6$\times10^{-3}$, maximum VMR of 1.6$\times10^{-2}$), and the range of values for Modern Earth (minimum VMR of 3.1$\times10^{-5}$, maximum VMR of 7.6$\times10^{-3}$). The Mid-Cretaceous, Neoproterozoic, and Modern values were taken from simulations of Earth epochs performed using the ROCKE-3D GCM and incorporating the SOCRATES radiation scheme \citep{rocke3d}, with all values other than \ce{CO2} and solar insolation left to modern Earth values (L. Sohl, priv. comm.). The particulars for each run are provided in Table~\ref{tab:rocke3d}.

\begin{deluxetable*}{ccccc}
    \tablehead{
    \colhead{Earth Epoch} & \colhead{Age} & \colhead{Solar Insolation} & \colhead{\ce{CO2}} & \colhead{Continent}
    }
    \startdata
        Neoproterozoic Slushball & ca. 715 Ma & 94\% modern solar insolation & 40 ppm & Sturtian Supercontinent \\
        Mid-Cretaceous Greenhouse & ca. 100 Ma & 99\% modern solar insolation & 2282 ppm & Mid-Cretaceous Continental \\
        Modern & pre-industrial (1850) & 100\% solar insolation (1361 W\ms) & 285 ppm & Modern Continental \\
    \enddata
    \caption{Details and parameter values for the ROCKE-3D simulations examining several epochs of Earth's history; values were provided by L. Sohl through private communication.}
    \label{tab:rocke3d}
\end{deluxetable*}

\section{Discussion}
\label{sec:discuss}

Understanding the impact of \ce{H2O} detectability across multiple selected bandpasses at varying wavelengths is illustrated in Figure~\ref{fig:shortest_detec_h2o}. This is crucial when looking into optimizing observations --- since next-generation telescope designs might prioritize shorter wavelengths based on photon flux or instrument efficiency, we can use these results to optimize observations by selecting a shorter bandpass that could meet different observational goals. As we can see in Figure~\ref{fig:shortest_detec_h2o}, \ce{O2} becomes strongly detectable at an SNR of 8. We can see that by carefully selecting the best bandpass, we can detect both \ce{O2} and \ce{H2O} at varying strengths, depending on the SNR needed. For example, if we wish to strongly detect both \ce{O2} and \ce{H2O} in the same bandpass, a slightly higher SNR is needed with the 0.74 {\microns} bandpass (SNR = 12) versus the 0.84 {\microns} bandpass (SNR = 11). 

There is also a trade-off between pure detectability versus an accurate constraint on abundance. If we purely want to confirm the presence of \ce{H2O}, we would prioritize the longer wavelengths close to 0.9 {\microns}, which are certain to yield a strong detection at mid to low SNRs for most abundance values; however, these bandpasses do not accurately retrieve the abundance values due to parameter degeneracies driven by a lack of spectral continuum. If we prioritize detecting an accurate abundance level for \ce{H2O} with a single bandpass measurement, we would need to utilize a shorter bandpass that requires higher SNRs and observation time. By prioritizing detectability above all else, we can confirm \ce{H2O} presence within more exo-atmospheres, and earmark more planets for follow-up observation at lower initial observation cost, which can then be observed for longer times and at higher SNRs (this will be explored further in a companion paper by \citet{stark23}). However, by prioritizing shorter wavelengths, the instrument capabilities are optimized and accurate abundance values retrieved, leading to better knowledge of the exoplanet composition (i.e.~slushball or water worlds) and more effectively designating interesting planets for follow-up observations of biosignatures. We note that a more thorough integration-time optimization can provide a firm single metric for evaluating these trades, but is immensely complex and model-dependent due to the combination of telescope and detector inputs, noise, resolution, etc., and thus is generally treated on a case by case basis for different mission designs. In this way, one can study the impact of specific parameters on the necessary integration time, such as in \citet{checlair21}, \citet{kopparapu21} and \citet{PSGbook}.

In Figure~\ref{fig:strong_quad_epochs} we examine the detectability of \ce{H2O} for several Earth epochs. We can see that detecting \ce{H2O} during a Neoproterozoic Slushball period would be difficult, requiring SNR = 7 in order to detect \ce{H2O} at 0.9 {\microns} or SNR = 9 at 0.85 {\microns}, with detection at shorter wavelengths unconstrained or improbable. Detecting a Mid-Cretaceous Greenhouse abundance would be relatively easy, with \ce{H2O} at multiple wavelengths detectable with mid to low SNRs; the caveat is that this is applicable only for the mid-to-high abundance values believed to be present for the Mid-Cretaceous Greenhouse. 
We can also see that the abundance values for Modern Earth cover a large range, encompassing all the same values of a Neoproterozoic Slushball and sharing a large portion of the Mid-Cretaceous Greenhouse. The Modern era values, however, do not climb to the same high abundances as the Mid-Cretaceous Greenhouse, and thus require an SNR of 10 to detect \ce{H2O} at all wavelengths of interest. \ce{H2O} at longer wavelengths, however, is accessible to be detected at an SNR of 4 at the larger abundance values. We also note that the ROCKE-3D simulations find a mean of 3.5$\times10^{-3}$ VMR (or 3500 ppm) for the Modern Earth Epoch, landing very close to our fiducial value of 3$\times10^{-3}$ VMR. All of the states discussed above, however, are to be used as examples of VMRs that may be possible for atmospheric conditions matching epochs of Earth's history, but we note that since there is little constraint on specific \ce{H2O} values over geologic time, one could not determine a specific epoch when a \ce{H2O} abundance is observed - the uncertainty and potential range of \ce{H2O} in these periods is large, thus at certain points a given abundance of \ce{H2O} could correspond to any of the three epochs we present. We present these values as a statement to the detectability of water vapor as a function of time using the best estimates from high-fidelity atmospheric models. 

As stated above, there have been similar prior works that investigated the relationship between SNR and detectability, in particular \citet{feng18} and \citet{damiano22}. Our overall retrieval parameterization structure is quite similar to that of F18, and in S23 we performed an apples-to-apples comparison with F18 to show that our results are largely equivalent; however, since F18 used the full visible wavelength range and we only use limited bandpasses in this work, we did not conduct a similar comparison. \citet{damiano22} also used much different wavelength ranges for their analyses, and also established the ability for their retrieval algorithm {\tt EXOREL} to determine the background gas of an atmosphere, whereas we define our background gas to be \ce{N2}. Even with these differences, our results for \ce{H2O} detectability agree with both \citet{damiano22} and \citet{feng18} that by an SNR of 10, at R = 140, firmly constraining the abundance for an Earth-twin atmosphere is very likely.

\section{Conclusions \& Future Works}
\label{sec:conc}

To summarize, SNR and detectability are intrinsically linked, and knowledge of the connection can drive efficient observing practices. By understanding the SNR requirements for detecting molecules of interest such as \ce{H2O}, and properly prioritizing the spectral bandpasses to optimize detectability of different atmospheric constituents, we can inform the best instrument designs and observing procedure. \ce{H2O} is most easily observable at longer wavelengths such as 0.9 {\microns}, however the stellar SED and instrument capabilities dictate that best observation occurs at shorter wavelengths, such as 0.74 {\microns}. Although the absorption is less deep in this wavelength range, one can also capture multiple molecular species such as \ce{O2}. Due to the lack of constraint on \ce{H2O} through time, observations of water vapor alone could not inform the epochs of an Earth-twin. However, other molecular species with geochemical markers could provide more depth of knowledge to potential Earth observations \citep{planavsky14}. 

In follow-up analyses, we will present a similar molecular abundance study for \ce{O2}, \ce{O3} and other potential molecular species that could be present in Earth-twin atmospheres, and expand on the trade space in choosing bandpasses for some or all of these species. 
We also note that there are a number of limitations to our model framework that we plan to improve on in subsequent work. We do not vary noise within the bandpasses as in real coronagraph observations (i.e.~prioritizing the bandpass center and increasing noise towards the bandpass edges), but rather assume a consistent noise throughout the bandpass. We use isotropic abundance profiles for the molecular species and cloudiness fraction, we have a limited selection of molecular species (i.e.~only \ce{H2O}, \ce{O2}, and \ce{O3}), and we also do not include any dependence on orbital phase. We also use units of geometric albedo for our spectral models; thus we cannot retrieve on radius due to the degeneracy described above in \S\ref{sec:method}. We intend to address these concerns in future works using new grids of forward models. We once again note that SNR results are subject to change with a new grid construction, due to upgrades in PSG and the addition of more parameters and molecules. This will allow us to represent the physical chemistry more accurately and study further molecular detections such as \ce{CH4} and \ce{CO2}. By improving and expanding the spectral grid, we will be able to expand our simulations of possible observations and establish best practices for exoEarth observations using next generation telescopes such as HWO. 
\\
\\
We thank the referee for their thoughtful review. N.~L. gratefully acknowledges financial support from an NSF GRFP. N.~L.~gratefully acknowledges Dr.~Joesph Weingartner for his support and editing, as well as Dr.~Ferah Munshi and Jordan Van Nest for their aid in plot coding. N.~L.~also gratefully acknowledges Greta Gerwig, Margot Robbie, Ryan Gosling, Emma Mackey, and Mattel Inc.{\texttrademark} for Barbie (doll, movie, and concept), for which this project is named after. This Barbie is an astrophysicist! The authors gratefully acknowledge conversations with Dr.~Giada Arney, Dr.~Michael Way, and Dr.~Linda Sohl on water vapor abundances in Earth's atmosphere through time. The authors also gratefully acknowledge conversation with Dr.~Chris Stark regarding exoEarth yields and instrument design. The authors would like to thank the Sellers Exoplanet Environments Collaboration (SEEC) and ExoSpec teams at NASA's Goddard Space Flight Center for their consistent support. MDH was supported by an appointment to the NASA Postdoctoral Program at the NASA Goddard Space Flight Center, administered by Oak Ridge Associated Universities under contract with NASA. 




\bibliography{main}

\end{document}